\documentclass{kapproc}
\usepackage{psfig}

\begin{document}

\articletitle[]{Optical Continuum Sources in \\
Gravitationally Lensed Quasars}

\author{Luis J. Goicoechea, David Alcalde}
\affil{Spanish Gravitational Lenses Group\footnote{Partial funding 
provided by the Spanish Dep. of Sci. and Tech. grant AYA2001-1647-C02.}}
\email{goicol@unican.es, dalcalde@ll.iac.es} 

\author{Vyacheslav N. Shalyapin}
\affil{Institute of Radio Astronomy, Ukraine}
\email{vshal@ira.kharkov.ua}

\begin{abstract}
We review some techniques to study the nature and size of the optical 
continuum sources in multiple QSOs. We focus on the source originating
the events with several months timescale (the rapid variability source)
as well as the source that is responsible for the non-variable background 
component (the background source). The techniques are used to study both 
the rapid variability source in Q0957+561 and the main (compact) 
background source in Q2237+0305.
\end{abstract}

\section{Techniques in the Time Domain}
Considering a simple flat universe (H$_0$ = 100 h km s$^{-1}$ Mpc$^{-1}$ 
and $\Omega_0$ = 1) and a far (z $\approx 1-2$) source with a radius R, the 
source's angular radius will be of $\theta$(arcsec) $\approx$ 1.7 10$^{-4}$ 
h R(pc). The optical continuum from QSOs could be mainly originated from an 
accretion disk around a 10$^8$ M$_\odot$ black hole. This standard scenario 
has a radius of about 0.01 pc, so $\theta \approx$ 10$^{-6}$ arcsec is the 
angular radius it subtends on the sky. The nuclear accretion disk is a good 
candidate to explain most of the non-variable background component, i.e., 
it may be the main background source, and the fluctuations on several 
months timescale (accretion disk instabilities). It is also evident that 
the violent variability may be originated in a circumnuclear stellar region, 
and the innermost stellar ring may be extended as far as $10^2-10^3$ pc. If 
R $\approx$ 10$^2$ pc then $\theta \approx$ 10$^{-2}$ arcsec. Angular 
resolutions less than a few hundredths of arcsec are a challenge for the 
current optical astronomy (including the Hubble Space Telescope). Therefore, 
instead of a direct mapping of distant nuclear and circumnuclear regions, we 
must do other experiments. In this contribution we review two experiments 
(in the time domain) to get valuable information on the properties of the 
optical continuum sources in gravitationally lensed quasars. 

Recently, Yonehara (1999) introduced a way to study the rapid variability 
source in multiple (lensed) QSOs. For a double-imaged QSO, with optical 
images A and B, there is a time delay between an event in image A and its 
twin event (a similar feature) in image B. If all features come from 
emitting points within the nuclear accretion disk, different delays for
different pairs of twin events cannot be resolved. However, if the violent 
events are generated in different points of the circumnuclear stellar 
region, it may be possible to find multiple delays (for layers at 10$^2$ pc, 
estimates of the maximum difference between delays, MDBD, are presented in 
Table 1). The timescale and the energy corresponding to the violent 
phenomena, the shape of the events and the results of chromatic tests 
(Collier 2001) are complementary data to decide on the physical processes 
taking place in the source.

\begin{table}[ht]
\caption[]{Estimates of the Maximum Difference Between Delays (MDBD) for 
Layers at 10$^2$ pc from the Centre of Several Lensed QSOs, and Comments
on their Possible Detection from Pairs of Twin Events with One--Day Error
Delays.}
\begin{tabular*}{\textwidth}{@{\extracolsep{\fill}}lcc}
\sphline
\it Lensed QSO &\it MDBD (days) &\it Detection\cr
\sphline
Q0909+532&8.7&Easy\cr
Q0921+4529&10.3&Easy\cr
Q0957+561&11.3&Easy\cr
Q1104--1805&13.9&Easy\cr
Q1600+434&3&Difficult\cr
\sphline
\end{tabular*}
\end{table}

On the other hand, from the analysis of a gravitational microlensing 
high-magnification event (HME) in an image of a multiple-imaged QSO, we can 
derive relevant results on the nature and physical parameters of the compact 
background source in the distant QSO (e.g., Wambsganss et al. 1990; Grieger 
et al. 1991; Rauch \& Blandford 1991; Webster et al. 1991). The simplest HMEs 
will result from the source either crossing a microcaustic or passing close 
to a microcusp, while more complex behaviours are expected in other cases: 
the source traveling through a network of microcaustics or the source 
traveling through a magnification region that quickly evolves as a 
consequence of stellar proper motions within the lens galaxy.

\section{Sources in Q0957+561 and Q2237+0305}
The recent monitoring of the Q0957+561 (double) and Q2237+0305 (quadruple) 
lensed quasars permitted to obtain accurate and well- sampled light curves 
(Kundi\'c et al. 1997; Serra-Ricart et al. 1999; Wo\'zniak et al. 2000; 
Alcalde et al. 2002). These modern records are basic tools to discuss on
the involved sources. The Q0957+561 brightness records include three pairs of 
twin violent events and do not show evidence for microlensing variability. 
Therefore, the Q0957+561 dataset is useful to analyze the rapid variability 
source at z = 1.41. Goicoechea (2002) found three different delays with MDBD 
$\approx$ 15 days (see the MDBDs in Table 1), suggesting that at least two 
pairs of events are generated in the circumnuclear stellar region. The energy 
and the timescale associated with the violent phenomena are consistent with 
the expected ones in a starburst scenario (e.g., Aretxaga et al. 1997). Other 
source models are also possible, in particular, a binary black hole. In this 
last picture, the two g-band pairs of events would come from the two 
separated accretion disks (see Collier 2001 for the possible origin of the main
g-band events).

The Q2237+0305 V-band light curves trace one HME in each of the brightest images 
A and C. In pioneer works with high-quality microlensing data, Yonehara (2001) 
analyzed the V-band HME in image C, whereas Shalyapin (2001) studied the two 
V-band HMEs. More recently, from a very detailed record of the peak in 
Q2237+0305A, Goicoechea et al. (2002) and Shalyapin et al. (2002) derived several 
results on the physical properties of the compact background source in Q2237+0305. 
If the microlensing peak is caused by a microcaustic crossing, they obtain that 
the standard accretion disk (e.g., Shakura \& Sunyaev 1973) fits both the V-band 
and R-band observations with reduced $\chi^2$ values very close to 1. Moreover, 
using the standard disk model and a robust upper limit on the transverse galactic 
velocity (Wyithe et al. 1999), the authors infer that 90 per cent of the V-band 
and R-band luminosities are emitted from a region with radial size less than 1.2 
10$^{-2}$ pc (= 3.7 10$^{16}$ cm, at 2$\sigma$ confidence level). Finally, 
assuming the existence of a nuclear standard disk, they obtain that the dark mass 
in the heart of the quasar must be larger than 10$^7$ M$_\odot$ and smaller than 6 
10$^8$ M$_\odot$ (see Figs. 1 and 2, which were derived from the {\it DSFor} 
parameters in Goicoechea et al. 2002 and reasonable values of the effective quasar 
velocity perpendicular to the microcaustic). 

\begin{figure}[ht]
\psfig{file=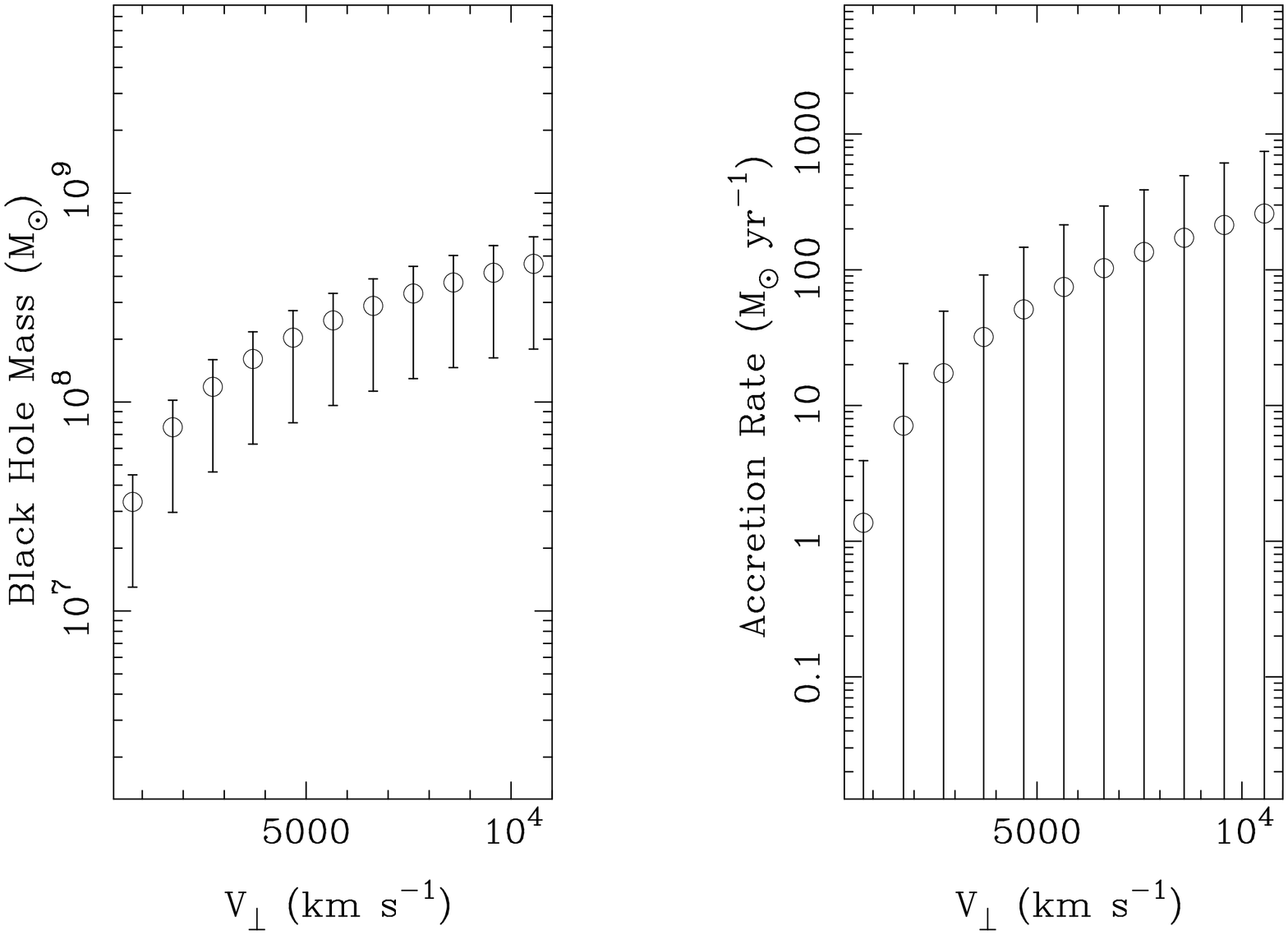,width=\textwidth,angle=-90}
\sidebyside
{\caption{Mass of the central black hole in Q2237+0305 (1$\sigma$ results).}}
{\caption{Mass accretion rate in the heart of Q2237+0305 (1$\sigma$ results).}}
\end{figure}

\end{document}